\documentstyle[preprint,prc,aps,psfig]{revtex}
   
  \def\vec#1{\mbox{\boldmath $#1$}}
  \def\gsl#1{\rlap{\slash}#1} 
  \def\Disc.{{\rm Im}}
  \def\Cont.{{\rm Re}}
  \def\<>#1{\langle{#1}\rangle} 
   
  \def\qq{\bar q q}
  \def\GG{{\alpha_s\over\pi}G^2}
  \def\mix{\bar q\sigma_{\mu\nu}G^{\mu\nu}q}
  \def\k{\gsl k} 
  \def\p{\gsl p} 
  \def\Im{{\rm Im}}
  \def\even{{\rm even}}
  \def\odd{{\rm odd}}
  \def\MN{{M}}
  \def\mpi{m_\pi}
  \def\MB{{M_B}}
  \def\MBS{{\MB^2}}
  \def\MBQ{{\MB^4}}
  \def\Eq.#1{Eq.~(\ref{#1})}

\title{A projected correlation function approach to the $\pi NN$ coupling constant in QCD sum rules}
\author{Yoshihiko Kondo\footnote{kondo@kokugakuin.ac.jp}}
\address{Kokugakuin University, Higashi, Shibuya, Tokyo 150-8440, Japan}
\author{Osamu Morimatsu\footnote{osamu.morimatsu@kek.jp}}
\address{Institute of Particle and Nuclear Studies, High Energy Accelerator Research Organization, Tukuba, Ibaragi 305-0801, Japan}

\begin{document}
\draft
\maketitle

\begin{abstract}
We propose a new approach to construct QCD sum rules for the $\pi NN$ coupling constant, $g$, starting from the vacuum-to-pion correlation function of the interpolating fields of two nucleons and taking its matrix element with respect to nucleon spinors.
The new approach with the projected correlation function is advantageous because even in the chiral limit the dispersion integral can be parametrized with well-defined physical parameters.
Another advantage of the new approach is that unwanted pole contribution is projected out.
Calculating the Wilson coefficients of the operator product expansion of the correlation function up to $O\left(\MB^{-4}\right)$ and $O\left(\mpi\right)$ where $\MB$ and $\mpi$ are the Borel mass and the pion mass, respectively,
we construct new QCD sum rules for the $\pi NN$ coupling constant from the projected correlation function with consistently including $O(\mpi)$ corrections.
By numerically analyzing the obtained four sum rules we identified the most prominent one.
After roughly estimating errors we obtaind, $g=10\pm3$, as a result of this sum rule, which is in reasonable agreement with the empirical value.
It is also found that  the $O(\mpi)$ correction is about 5\%.
\end{abstract}


\newpage

\section{INTRODUCTION}

Within the framework of the QCD sum rule proposed by Shifman, Vainshtein and Zakharov~[\ref{SVZ}],
much work has been done for the meson-baryon-baryon coupling constants ~[\ref{RRY}-\ref{DKO}]. 
Sum rules have been constructed in two formulations:
one is to start from the vacuum-to-vacuum matrix element of the correlation function of the interpolating fields of two baryons and one meson and the other is to start from the vacuum-to-pion matrix element of the correlation function of the interpolating fields of two baryons.
In the pionnering work by Reinders, Rubinstein and Yazaki the former approach was taken but in the following works including the one by the same authors the latter approach was employed since some difficulties seem to exist in the former approach~[\ref{Maltman}].

Therefore, we discuss only the formulation with the vacuum-to-pion matrix element of the correlation function of the interpolating fields of two nucleons:
\begin{eqnarray}\label{Pi}
\Pi(p,k)=
-i\int d^4xe^{ipx}\langle 0|T[\eta(x)\bar\eta(0)]|\pi(k)\rangle,
\end{eqnarray}
where $\eta(x)$ is the interpolating field for the nucleon, $|\pi(k)\rangle$ is the pion state with momentum $k$ normalized as $\langle\pi(k')|\pi(k)\rangle=2k_0(2\pi)^3\delta^3(\vec k'-\vec k)$ and the isospin is neglected for simplicity.
The correlation function can be split into four Dirac structures,
$i\gamma_5$, $i\gamma_5\p$, $i\gamma_5\k$ and $\gamma_5\sigma_{\mu\nu}p^\mu k^\nu$.
The coefficients of the Dirac structures are functions of Lorentz invariants, $p^2$ and $pk$, and are called invariant correlation functions.
Reinders, Rubinstein and Yazaki constructed a sum rule for the pion-nucleon-nucleon ($\pi NN$) coupling constant in the chiral limit with vanishing pion momentum from the invariant correlation function with the Dirac structure $i\gamma_5$ in the leading order of the operator product expansion (OPE).
They showed that if the sum rule is divided by the odd-dimensional sum rule for the nucleon mass, the result is consistent with the Goldberger-Treiman relation with $g_A=1$, where $g_A$ is the axial charge of the nucleon~[\ref{RRY}].
Shiomi and Hatsuda improved the sum rule by taking into account higher dimensional terms and $\alpha_s$ corrections of the OPE and also the continuum contributions~[\ref{S&H}].
They showed that even after these corrections are taken into account, the Goldberger-Treiman relation with $g_A=1$ holds as long as the same continuum threshold is taken in the sum rule for the $\pi NN$ coupling constant and the odd-dimensional sum rule for the nucleon mass.
In Ref.~[\ref{B&K}] Birse and Krippa pointed out that in the chiral limit the vacuum-to-pion correlation function is obtained just by chirally rotating the vacuum-to-vacuum correlation function and therefore it is obvious that the ratio of the two sum rules is consistent with the Goldberger-Treiman relation, which is just a consequence of the chiral symmetry.
Then, they tried to obtain a new sum rule for the $\pi NN$ coupling constant which is not just a consequence of the chiral symmetry.
They considered the invariant correlation function with the structure $i\gamma_5\gsl k$ and obtained a sum rule by taking the limit, $k=0$, after removing $i\gamma_5\gsl k$.
In all of these works the sum rule is constructed by making an \lq\lq ansatz \rq\rq for the absorptive part of the correlation function based on the effective lagrangian with the pseudoscalar coupling scheme.
In Ref.~[\ref{KLO}] Kim, Lee and Oka examined how the choices of the effective lagrangian, i.e. the pseudoscalar and pseudovector coupling schemes make differences in the sum rules for all the four Dirac structures.
They concluded that only the invariant correlation function with the Dirac structure $\gamma_5\sigma_{\mu\nu}p^\mu k^\nu$, is independent of the two coupling schemes and obtained a sum rule by taking the limit, $k=0$, after removing $\gamma_5\sigma_{\mu\nu}p^\mu k^\nu$.
They also claimed that their sum rule suffers from less uncertainties due to QCD parameters.

In a recent work, we have studied in detail the physical content of the invariant correlation functions without referring to the effective lagrangian~[\ref{recentKM}].
We have shown that the coefficients of the double poles are proportional to the $\pi NN$ coupling constant
but that the coefficients of the single poles are not determined by the $\pi NN$ coupling constant.
In the chiral limit the double-pole terms survive only for the Dirac structures $i\gamma_5\gsl k$ and $\gamma_5\sigma_{\mu\nu}p^\mu k^\nu$.
For these structures the single-pole terms as well as the continuum terms turn out to be ill defined in the dispersion integral.
Therefore, the use of naive QCD sum rules for these structures in the chiral limit is not justified.

In this paper we propose a new approach, a projrcted correlation function approach.
We investigate the QCD sum rule for the $\pi NN$ coupling constant derived from a projrcted correlation function by applying the procedure of Ref.~[\ref{K&M}].
An advantage of the new approach is that the discontinuity of the projected correlation function does not include ill-defined terms even in the chiral limit in contrast to the previous approaches.
Another point is that we obtain sum rules not only in the chiral limit but also with consistently including $O(\mpi)$ corrections.

This paper is organized as follows.
In Sec. II, we summarize the physical content of the correlation function without referring to the effective lagrangian.
We introduce the projected correlation function and study its physical content in Sec. III.
In Sec. IV we construct QCD sum rules for the projected correlation function and investigate the numerical results of the obtained sum rules.
Section V is devoted to summary.

\section{PHYSICAL CONTENT OF THE CORRELATION FUNCTION}

In this section, we summarize the physical content of the correlation function defined by \Eq.{Pi} without referring to the effective theory~[\ref{recentKM}].
The correlation function $\Pi(p,k)$ in \Eq.{Pi} has a pole at $p^2=\MN^2$ and $(p-k)^2=\MN^2$ where $p$ and $p-k$, respectively, become on-shell momenta for the nucleon where $\MN$ denotes the nucleon mass.
The $\pi NN$ coupling constant, $g$, is defined through the coefficient of the pole as
\begin{eqnarray}\label{gpiNN}
\bar u(\vec pr)(\p-\MN)\Pi(p,k)(\p-\k-\MN)u(\vec qs)|_{p^2=\MN^2,(p-k)^2=\MN^2}=-i\lambda^2g\bar u(\vec pr)\gamma_5u(\vec qs),
\end{eqnarray}
where $q=p-k$, $u(\vec pr)$ is a Dirac spinor with momentum $p$, spin $r$ and is normalized as $\bar u(\vec pr)u(\vec pr)=2\MN$.
$\lambda$ is the coupling strength of the interpolating field to the nucleon.

In addition to the pole singularity, $\Pi(p,k)$ has a branch cut singularity starting from the threshold of the pion-nucleon channel, $p^2=(\MN+\mpi)^2$, to infinity where $\mpi$ is the pion mass.
We restrict ourselves to the positive energy region, $p_0>0$, for simplicity.

In order to classify the singularity of $\Pi(p,k)$,
we define the $\pi NN$ vertex function, $\Gamma(p,q',k')$ $(p=q'+k')$, and the pion-nucleon T-matrix, $T(q',k',q,k)$ $(q'+k'=q+k)$, by
\begin{eqnarray}
&&\Gamma(p,q',k')=(\p-\MN)\left[-i\int d^4xe^{ipx}\langle 0|T[\eta(x)\bar\eta(0)]|\pi(k')\rangle\right](\gsl q'-\MN),\cr
&&T(q',k',q,k)=(\gsl q'-\MN)\left[-i\int d^4xe^{iq'x}\langle\pi(k')|T[\eta(x)\bar\eta(0)]|\pi(k)\rangle\right](\gsl q-\MN).
\end{eqnarray}
The correlation function is related to the vertex function as
\begin{eqnarray*}
\Pi(p,k)={\p+\MN \over p^2-\MN^2}\Gamma(p,q,k){\p-\k+\MN\over
(p-k)^2-\MN^2}.
\end{eqnarray*}

Let us regard $\Pi(p,k)$ as a function of the center-of-mass energy, i.e. $p_0$ in the frame $\vec p=0$, and consider the discontinuity of the correlation function.
We adopt the following notation for the dispersive (continuous) part and the absorptive (discontinuous) part, respectively:
\begin{eqnarray}
&&\Cont.F(p)\equiv{1\over2}\left[F(p)|_{p^0=p^0+i\eta}+F(p)|_{p^0=p^0-i\eta}\right],\cr
&&\Disc.F(p)\equiv{1\over2i}\left[F(p)|_{p^0=p^0+i\eta}-F(p)|_{p^0=p^0-i\eta}\right].
\end{eqnarray}
The absorptive part of the correlation function, ${\rm Im}\Pi(p,k)$, can be written as
\begin{eqnarray}
{\rm Im}\Pi(p,k)
=(\p+\MN)\bigg\{&&{\rm Im}{1 \over (p^2-\MN^2)((p-k)^2-\MN^2)}{\rm Re}\Gamma(p,q,k)\cr
& &+{\rm Re}{1 \over (p^2-\MN^2)((p-k)^2-\MN^2)}{\rm Im}\Gamma(p,q,k)\bigg\}
(\p-\k+\MN).
\end{eqnarray}
The first term represents the pole contribution and the second term represents the continuum contribution.

When the center-of-mass energy, $p_0$, is above the threshold of the pion-nuclon channel, $\MN+m_{\pi}$, but below that of the next channel, only the pion-nucleon channel contributes in the intermediate states and the absorptive part of the vertex function, ${\rm Im}\Gamma(p,q,k)$, is given by
\begin{eqnarray}\label{O_T}
\Disc.\Gamma(p,q,k)
&=&-{\pi\over4}{t\over p_0}\int {d\Omega'\over(2\pi)^3}
{\Gamma(p,q',k')}(\gsl q'+\MN){T^*(q',k',q,k)},
\end{eqnarray}
where $t$ is defined through $p_0=\sqrt{\MN^2+t^2}+\sqrt{m_{\pi}^2+t^2}$.
$q'$ and $k'$ are the on-shell momenta for the nucleon and the pion, respectively.

From \Eq.{O_T}, one sees that in the vicinity of the pion-nucleon threshold $\Gamma$ can be expanded as
\begin{eqnarray}\label{ImVF}
\begin{array}{l}
  {\rm Im}\Gamma(p,q,k)=\left\{
    \begin{array}{l l}
      0&\quad(p_0< \MN+\mpi)\cr
      G_1t
      + O(t^3)&\quad(p_0> \MN+\mpi)
    \end{array}
\right. ,\cr
  {\rm Re}\Gamma(p,q,k)=\left\{
    \begin{array}{l l}
      G_0+G_1\tau+O(\tau^3)&\quad(p_0< \MN+\mpi)\cr
      G_0+O(t^2)&\quad(p_0>  \MN+\mpi)
    \end{array}
\right. ,
\end{array}
\end{eqnarray}
where $\tau=it$.
In \Eq.{ImVF} the behaviour of the dispersive part is determined from that of the absorptive part by analytic continuation. 
From \Eq.{ImVF} one sees that
\begin{eqnarray}\label{ReVF}
\left.{\partial\over\partial p_0}{\rm Re}\Gamma(p,q,k)\right|_{p_0=\MN+\mpi-\epsilon}&=&\left\{
    \begin{array}{l l}
      \infty&\quad(\mpi\not=0)\cr
      \hbox{finite constant}&\quad(\mpi=0)
    \end{array}\right. ,\cr
\left.{\partial\over\partial p_0}
{\rm Re}\Gamma(p,q,k)\right|_{p_0=\MN+\mpi+\epsilon}&=&0.
\end{eqnarray}
We found in Ref.[\ref{recentKM}] that in the chiral limit the double-pole terms survive only for the invariant correlation function with the Dirac structures $i\gamma_5\gsl k$ and $\gamma_5\sigma_{\mu\nu}p^\mu k^\nu$.
The coefficients of the single poles for the structures $i\gamma_5\gsl k$ and $\gamma_5\sigma_{\mu\nu}p^\mu k^\nu$ involve the derivatives of the vertex functions with respect to $p^2$ and are therefore ill defined due to \Eq.{ReVF}.
Therefore, the use of naive QCD sum rules for these structures in the chiral limit is not justified.

A reasonable procedure to make an ansatz for the absorptive part of the invariant correlation function to be used in the sum rule would be first to transform it, for instance by multiplying by $p^2-\MN^2$, to a form which does not include ill-defined terms. 
However as pionted out in Refs.[\ref{B&K},\ref{KLO}] one is unable to find a reasonable result using this proceedure.
In this paper, we will propose a different way to construct a sum rule.

\section{PHYSICAL CONTENT OF THE PROJECTED CORRELATION FUNCTION}
We define the projected correlation function, $\pi NN$ vertex function and pion-nucleon T-matrix by
\begin{eqnarray}\label{Pi+}
&&\Pi_+(pr,qs,k)=\bar u(\vec pr)\gamma_0\Pi(p,k)\gamma_0u(\vec qs),\cr
&&\Gamma_+(pr,qs,k)=\bar u(\vec pr)\Gamma(p,q,k)u(\vec qs),\cr
&&T_+(qs,k,q's',k')=\bar u(\vec qs)T(q,k,q',k')u(\vec q's').
\end{eqnarray}
The projected vertex function is related to the projected correlation function as
\begin{eqnarray}\label{Vertex}
\Gamma_+(pr,qs,k)=(p_0-\MN)(p_0-E_q-\omega_k)\Pi_+(pr,qs,k),
\end{eqnarray}
where $E_q=\sqrt{\MN^2+\vec q^2}$ and $\omega_k=\sqrt{\mpi^2+\vec k^2}$.
It should be noted that $\Pi_+(pr,qs,k)$ has poles at $p_0=\MN$ and $p_0=E_q+\omega_k$ but not at $p_0=-\MN$ and $p_0=-E_q+\omega_k$.
Then the absorptive part of the projected correlation function, $\Disc.\Pi_+$, can be written as
\begin{eqnarray}\label{DiscPi}
\Disc.\Pi_+(pr,qs,k)
=&&\pi\delta(p_0-\MN){\Cont.\Gamma_+(pr,qs,k)\over E_q+\omega_k-\MN}-\pi\delta(p_0-E_q-\omega_k)
{\Cont.\Gamma_+(pr,qs,k)\over E_q+\omega_k-\MN}
\cr && +\Cont.{1\over (p_0-\MN)(p_0-E_q-\omega_k)}\Disc.\Gamma_+(pr,qs,k).
\end{eqnarray}
Let us now consider the last term of \Eq.{DiscPi}.
When the center-of-mass energy, $p_0$, is above the threshold of the pion-nucleon channel, $\MN+m_{\pi}$, but below that of the next channel, $\Disc.\Gamma_+(pr,qs,k)$ is given by
\begin{eqnarray}\label{ImGP}
\Disc.\Gamma_+(pr,qs,k)
&=&-{\pi\over4}{t\over p_0}\sum_{s'}\int {d\Omega\over(2\pi)^3}
{\Gamma_+(pr,q's',k')}{T_+^*(q's',k',qs,k)}\cr
&=&-{\pi\over4}{t\over p_0}\lambda^2g(p_0,t^2)\int {d\Omega\over(2\pi)^3}
\bar u(\vec pr)i\gamma_5(\gsl q'+\MN)T^*(q'k',q,k)u(\vec qs),
\end{eqnarray}
where $q'$ and $k'$ are on-shell momenta for the nucleon and the pion, respectively, and $g(p_0,t^2)$ is defined by
\begin{eqnarray}
\Gamma_+(pr,q's',k')=\lambda^2\bar u(\vec pr)i\gamma_5 u(\vec q's')g(p_0,t^2).
\end{eqnarray}
Since $\bar u(\vec pr)i\gamma_5(\gsl q'+\MN)\rightarrow\bar u(\vec pr)i\gamma_5(-t^2/2\MN-\gamma_iq'_i)\rightarrow0$ as $p_0\rightarrow \MN+m$, $\Gamma_+$ behaves in the vicinity of the pion-nucleon threshold as
\begin{eqnarray}\label{ImReGP}
  &&{\rm Im}\Gamma_+(pr,qs,k)=\left\{
    \begin{array}{l l}
      0&\quad(p_0< \MN+\mpi)\cr
      O(t^3)&\quad(p_0> \MN+\mpi)
    \end{array}
\right.,\cr
  &&{\rm Re}\Gamma_+(pr,qs,k)=\left\{
    \begin{array}{l l}
      G_{+}+O(\tau^3)&\quad(p_0< \MN+\mpi)\cr
      G_{+}+O(t^2)&\quad(p_0>  \MN+\mpi)
    \end{array}
\right.,
\end{eqnarray}
which is in contrast to \Eq.{ImVF}.

Here we should discuss the behaviour of $\Im \Pi_+$ in the chiral limit with $ k=0$:
We obtain
\begin{eqnarray}\label{ImPi}
\Disc.\Pi_+(pr,qs,k)
=&&-\pi\delta'(p_0-\MN)\Cont.\Gamma_+(pr,qs,k)
+\pi\delta(p_0-\MN){\partial\over\partial p_0}\Cont.\Gamma_+(pr,qs,k)
\cr && +{\rm Re}{1\over(p_0-\MN)^2}\Disc.\Gamma_+(pr,qs,k).
\end{eqnarray}
The coefficient of the single pole term vanishes since ${\partial\over\partial p_0}\Cont.\Gamma_+$ behave as $(p_0-\MN)^2$ for $p_0<\MN$ and $p_0-\MN$ for $p_0>\MN$. 
The last term behaves as $\theta(p_0-\MN)(p_0-\MN)$ since $\Disc.\Gamma_+$ behaves as $\theta(p_0-\MN)(p_0-\MN)^3$.
Therefore, the second term and third term on the right-hand side of \Eq.{ImPi} are well defined in the dispersion integral.


\section{CONSTRUCTION OF SUM RULES}
For the sake of completeness we first summarize the derivation of the Borel sum rule for the projected correlation function.
The dispersion relation for the projected correlation function in the variable $p_0$ is
\begin{eqnarray}\label{Dispersion}
\Pi_+(p_0,\vec p)&=&-{1\over\pi}\int dp_0'
{\Disc.\Pi_+(p_0',\vec p)\over p_0-p_0'+i\eta}.
\end{eqnarray}
By splitting the projected correlation function into the even part, $\Pi_{+\even}={1\over2}[\Pi(p_0,\vec p)+\Pi(-p_0,\vec p)]$, and odd one, $\Pi_{+\odd}={1\over2p_0}[\Pi(p_0,\vec p)-\Pi(-p_0,\vec p)]$, we rewrite \Eq.{Dispersion} in terms of the even and odd parts as
\begin{eqnarray}\label{Dispersion2}
\Pi_{+\even}(p_0^2,\vec p)&=&-{1\over\pi}\int dp_0'
{p_0'\over p_0^2-p_0'^2}\Disc.\Pi_+(p'_0,\vec p),
\cr
\Pi_{+\odd}(p_0^2,\vec p)&=&-{1\over\pi}\int dp_0'
{1\over p_0^2-p_0'^2}\Disc.\Pi_+(p'_0,\vec p).
\end{eqnarray}
Applying the Borel transformation,
\begin{eqnarray*}
       L_B\equiv
       \lim_{{n\rightarrow\infty \atop -p_0^2\rightarrow\infty}
       \atop -p_0^2/n = \MBS}
       {(p_0^2)^n\over(n-1)!}\left(-{d\over dp_0^2}\right)^n ,
\end{eqnarray*}
where $\MB$ is called the Borel mass, we get
\begin{eqnarray}\label{BSR}
L_B[\Pi_{+\even}(p_0^2,\vec p)]
&=&{1\over\pi}\int dp_0'{p_0'\over\MBS}\exp\left(-{{p_0'}^2\over\MBS}\right)
\Disc.\Pi_+(p'_0,\vec p),
\cr
L_B[\Pi_{+\odd}(p_0^2,\vec p)]
&=&{1\over\pi}\int dp_0'{1\over\MBS}\exp\left(-{{p_0'}^2\over\MBS}\right)
\Disc.\Pi_+(p'_0,\vec p).
\end{eqnarray}
These equations with the correlation functions on the left-hand side evaluated by the OPE are the Borel sum rules.


Now, we calculate the OPE of the two-point correlation function.
We consider the following correlation function with charged pion,
\begin{eqnarray}\label{Pi_pi+}
\Pi(p,k)
&=&-i\int d^4xe^{ipx}\langle0|T[\eta_p(x)\bar\eta_n(0)]|\pi^+(k)\rangle.
\end{eqnarray}
with the interpolating field for the proton and the neutron as~[\ref{Ioffe}]
\begin{eqnarray*}
\eta_p&=&\epsilon_{abc}[u_a^TC\gamma_\mu u_b]\gamma_5\gamma^\mu d_c,\cr
\eta_n&=&-\epsilon_{abc}[d_a^TC\gamma_\mu d_b]\gamma_5\gamma^\mu u_c,
\end{eqnarray*}
where $u$ and $d$ are the up-quark field and the down-quark field, respectively, $C$ is the charge conjugation operator and $a$, $b$ and $c$ are color indices.
The Wilson coefficients for eq.(\ref{Pi_pi+}) have been calculated in Ref.~[\ref{RRY}-\ref{KLO}].
$O(k^0)$ terms for the structure $i\gamma_5$, $i\gamma_5\gsl k$ and $\gamma_5\sigma_{\mu\nu}k^\mu p^\nu$ are obtained in Ref.~[\ref{RRY},\ref{S&H}], [\ref{B&K}] and [\ref{KLO}], respectively.
Ref~[\ref{KLO}] also gives $O(k)$ terms for the structure $i\gamma_5$.
Here, we also need $O(k)$ terms for the structure $i\gamma_5\gsl p$, $i\gamma_5\gsl k$ and $\gamma_5\sigma_{\mu\nu}k^\mu p^\nu$.
We calculate the Wilson coefficients of the short-distance expansion in two steps:
we perform the light-cone expansion of the correlation function first and the short-distance expansion of the light-cone operators second.
The reason for doing this is to use the parameterization of the vacuum-to-pion matrix elements of the light-cone operators given in Ref.[\ref{Belyaev}],
\begin{eqnarray}\label{ME}
\langle0|\bar d(0)i\gamma_5u(x)|\pi^+(k)\rangle
&=&\sqrt{2}{f_\pi \mpi^2\over m_u+m_d}\left(1-{i\over2}kx+O\left((kx)^2\right)\right),
\cr 
\langle0|\bar d(0)\gamma_5\sigma_{\mu\nu}u(x)|\pi^+(k)\rangle
&=&\sqrt{2}i(k_\mu x_\nu-k_\nu x_\nu){f_\pi \mpi^2\over6(m_u+m_d)}\left(1-{i\over2}kx+O\left((kx)^2\right)\right),
\cr 
\langle0|\bar d(0)\gamma_5\gamma_\mu u(x)|\pi^+(k)\rangle
&=&\sqrt{2}ik_\mu f_\pi\left(1-{i\over2}kx+O\left((kx)^2\right)\right)
\cr& &+\sqrt{2}ik_\mu x^2{5f_\pi\delta^2\over36}\left(1-{i\over2}kx+O\left((kx)^2\right)\right)
\cr& &-\sqrt{2}i x_\mu kx{f_\pi\delta^2\over18}\left(1-{i\over2}kx+O\left((kx)^2\right)\right),
\cr 
\langle0|\bar d(0)g_s\tilde G_{\alpha\beta}(0)\gamma_\mu u(x)|\pi^+(k)\rangle
&=&-\sqrt{2}i(k_\beta g_{\alpha\mu}-k_\alpha g_{\beta\mu})f_\pi\delta^2\cr
& &\quad\times\left({1\over3}-ikx{7+6\epsilon\over84}+O\left((kx)^2\right)\right)
\cr& &+\sqrt{2}(k_\alpha x_\beta-k_\beta x_\alpha)k_\mu f_\pi\delta^2\cr
& &\quad\times\left({7-6\epsilon\over252}
-ikx{4-3\epsilon\over504}+O\left((kx)^2\right)\right).
\end{eqnarray}
$f_\pi$ is the pion decay constant, $m_u$ ($m_d$) is the up(down)-quark mass, $G_{\rho\sigma}$ is the gluon field tensor and $\tilde G_{\alpha\beta}={1\over2}\epsilon_{\alpha\beta\mu\nu}G^{\mu\nu}$.
$\delta^2$ is defined through $g_s\langle0|\bar d\tilde G_{\mu\nu}\gamma^\nu u|\pi^+(k)\rangle=\sqrt{2}i f_\pi \delta^2 k_\mu$ where $g_s$ is the coupling constant of the QCD.
$\epsilon$ is a parameter associated with the deviation of twist-4 pion wave functions from their asymptotic forms [\ref{Braun}].

Going through the above two steps, we obtain the OPE of the correlation function as
\begin{eqnarray}\label{OPE}
\Pi(p,k)
&=&-{\sqrt{2}\over4\pi^2}\Bigg\{
i\gamma_5\Bigg[ p^2\ln(-p^2){\<>{\qq}\over f_\pi}
+{1\over p^2}\left( -{\pi^2\over6}{\<>{\qq}
\<>{\GG}\over f_\pi} \right) \Bigg]
\cr& &\qquad+i\gamma_5\k\Bigg[ p^2\ln(-p^2)2f_\pi+\ln(-p^2)2f_\pi\delta^2
+{1\over p^2}\left({8\pi^2\over9}{\<>{\qq}^2\over f_\pi}
+{\pi^2\over3}f_\pi\<>{\GG}\right)
\cr& &\qquad\qquad +{1\over p^4}\left( -{\pi^2\over9}
{\<>{\qq} g_s\<>{\mix}
\over f_\pi} \right) \Bigg]
\cr& &\qquad+\gamma_5\sigma_{\mu\nu}p^\mu k^\nu\Bigg[
\ln(-p^2){\<>{\qq}\over3f_\pi}
+{1\over p^2}{16\pi^2\over3}f_\pi\<>{\qq}
\cr& &\qquad\qquad+{1\over p^4}\left(
{2\pi^2\over3}f_\pi g_s\<>{\mix}
-{\pi^2\over54}{\<>{\qq}\<>{\GG}
\over f_\pi}
-{104\pi^2\over27}f_\pi\delta^2\<>{\qq}\right) \Bigg] 
\cr& &\qquad+i\gamma_5 p k\Bigg[
\ln(-p^2)\left(-{\<>{\qq}\over f_\pi}\right)
+{1\over p^4}\left(-{\pi^2\over6}
{\<>{\qq}\<>{\GG}\over f_\pi}
\right)
\Bigg]
\cr& &\qquad+i\gamma_5\p p k\Bigg[
{1\over p^2}\left(-{4\over9}f_\pi\delta^2\right)
+{1\over p^4}{16\pi^2\over9}{\<>{\qq}^2\over f_\pi}
+{1\over p^6}{4\pi^2\over9}{\<>{\qq}\over f_\pi} 
g_s\<>{\mix} \Bigg]
\cr& &\qquad+i\gamma_5\k p k\Bigg[
\ln(-p^2)(-2f_\pi)+{1\over p^2}\left(-{16\over9}f_\pi\delta^2 \right)
+{1\over p^4}{\pi^2\over3}f_\pi\<>{\GG}
 \Bigg] 
\cr& &\qquad+\gamma_5\sigma_{\mu\nu}p^\mu k^\nu p k\Bigg[
{1\over p^2}\left(-{1\over3}{\<>{\qq}\over f_\pi}\right)
+{1\over p^4}{16\pi^2\over3}f_\pi\<>{\qq}
+{1\over p^6}\Bigg({4\pi^2\over3}f_\pi g_s\<>{\mix}
\cr& &\qquad\qquad
-{448+32\epsilon\over63}{\pi^2}f_\pi\delta^2\<>{\qq}
-{\pi^2\over9}{\<>{\qq}\over f_\pi}\<>{\GG} \Bigg)
 \Bigg]\Bigg\},
\end{eqnarray}
where $\langle\hat O\rangle$ denotes the vacuum-to-vacuum matrix element of the operator $\hat O$, $\langle\bar qq\rangle=\langle\bar uu\rangle=\langle\bar dd\rangle$ and $G^2=2{\rm tr}\left(G_{\mu\nu}G^{\mu\nu}\right)$.

Some comments are in order here.
The matrix elements of the four-quark operators and the mixed quark-gluon operators are approximated to factorize.
\Eq.{OPE} is expressed in terms of the quark condensate, $\langle\bar qq\rangle$, by the use of the Gell-Mann-Oakes-Renner relation, $f_\pi^2 \mpi^2=-(m_u+m_d)\langle\bar qq\rangle$.
Though one might think that the terms with $i\gamma_5\k$ and $\gamma_5\sigma_{\mu\nu}p^\mu k^\nu$ are of higher order in $k$ than those with $i\gamma_5$ and $i\gamma_5\p$, the former and the latter are of the same order when sandwiched by $\bar u(\vec pr)\gamma_0$ and $\gamma_0u(\vec qs)$ in the center-of-mass frame:
\begin{eqnarray*}
  \bar u(\vec pr)\gamma_0i\gamma_5\k\gamma_0u(\vec qs)
&=&(E_{q}+\MN-\omega_k)\bar u(\vec pr)\gamma_0i\gamma_5\gamma_0u(\vec qs)\cr
&=&{\omega_k-E_{q}-\MN\over p_0}\bar u(\vec pr)\gamma_0i\gamma_5\p\gamma_0u(\vec qs),
\end{eqnarray*}
\begin{eqnarray*}
  \bar u(\vec pr)\gamma_0\gamma_5\sigma_{\mu\nu}p^\mu k^\nu\gamma_0u(\vec qs)
&=&-p_0(E_{q}+\MN)\bar u(\vec pr)\gamma_0i\gamma_5\gamma_0u(\vec qs)\cr
&=&(E_{q}+\MN)\bar u(\vec pr)\gamma_0i\gamma_5\p\gamma_0u(\vec qs).
\end{eqnarray*}
Thus, in \Eq.{OPE} we showed the terms which survive up to $O\left(p_0^{-4}\right)$ and $O\left(\omega_k\right)$ after projection, and therefore the dimension of the operator does not play the role of the expansion parameter.


Next, we express the right-hand side of the Borel sum rule, \Eq.{BSR}, by physical quantities.
We would like to make an ansatz for the absorptive part of the projected correlation function of \Eq.{Pi+} to be used in the sum rule.
Since \Eq.{ImPi} does not include irregular terms, 
we make the following ansatz for the absorptive part of the projected correlation function for the $\pi^+$-neutron-proton vertex:
\begin{eqnarray}\label{ImPart}
\Disc.\Pi_+(pr,qs,k)
=&&\pi\delta(p_0-\MN){\Cont.\Gamma_+(pr,qs,k)\over E_q+\omega_k-\MN}-\pi\delta(p_0-E_q-\omega_k)
{\Cont.\Gamma_+(pr,qs,k)\over E_q+\omega_k-\MN}\cr
&&+\left[\theta(p_0-\omega_\pi)+\theta(-p_0-\omega_\pi)\right]\Disc.\Pi_+^{\rm OPE}(pr,qs,k)\cr
=&&\delta(p_0-\MN)\bar u(\vec pr)i\gamma_5u(\vec qs)\sqrt 2\pi\lambda^2{g(M,\vec k^2)\over E_q+\omega_k-\MN}\cr
&&-\delta(p_0-E_q-\omega_k)\bar u(\vec pr)i\gamma_5u(\vec qs)\sqrt 2\pi\lambda^2{g(E_q+\omega_k,\vec k^2)\over E_q+\omega_k-\MN}\cr
&&+\left[\theta(p_0-\omega_\pi)+\theta(-p_0-\omega_\pi)\right]\Disc.\Pi_+^{\rm OPE}(pr,qs,k),
\end{eqnarray}
where $\omega_\pi$ is the effective continuum threshold of $\pi N$ or $\pi\bar N$ channel and $\sqrt{2}$ on the right-hand side is the isospin factor.

As is explained in \Eq.{gpiNN}, the $\pi NN$ coupling constant is defined at $p^2=\MN^2$, $(p-k)^2=\MN^2$, i.e.,
\begin{eqnarray}
g=g\left(p_0=\MN,{\vec k}^2=-\mpi^2+{\mpi^4\over 4\MN^2}\right),
\end{eqnarray}
while $g(M,\vec k^2)$ and $g(E_q+\omega_k,\vec k^2)$ in \Eq.{ImPart} refer to different kinematical points.
However, if one expands them in $\mpi$ taking $\vec k=0$ one finds:
\begin{eqnarray*}\label{g(\MN)-g}
&&g(\MN,0)=g+O\left(\mpi^2\right),\cr
&&g(\MN+\mpi,0)=g(\MN,0)+\mpi\left.{\partial g(p_0,0)\over\partial p_0}\right|_{p_0=\MN}+O\left(\mpi^2\right)=g+g'\mpi+O\left(\mpi^2\right).
\end{eqnarray*}
Therefore, if we are interested only in the result of $O(\mpi)$, the difference of $g(\MN,0)$ and $g$ can be ignored.
Then, substituting \Eq.{ImPart} into the right-han side of \Eq.{BSR} and expanding in $O(\mpi)$ we obtain
\begin{eqnarray}\label{DisInt}
&&{1\over\pi}\int dp_0'{p_0'\over\MBS}\exp\left(-{{p_0'}^2\over\MBS}\right)
\Disc.\Pi_+(p'r,qs,k)\cr
&=&-\bar u(\vec pr)i\gamma_5u(\vec qs){2\MN\over\MBS}\left(g-{g'\MN+g\over2\MN^2}\MBS\right){\sqrt 2\lambda^2\over \MBS}\exp\left(-{\MN^2\over \MBS}\right)+({\rm Cont.})
\cr
&&{1\over\pi}\int dp_0'{1\over\MBS}\exp\left(-{{p_0'}^2\over\MBS}\right)
\Disc.\Pi_+(p'r,qs,k)\cr
&=&-\bar u(\vec pr)i\gamma_5u(\vec qs){2\MN\over\MBS}\left(g-{g'\over 2\MN}\MBS\right){\sqrt 2\lambda^2\over \MBS}\exp\left(-{\MN^2\over \MBS}\right)+({\rm Cont.}),
\end{eqnarray}
where $O(\mpi^2)$ terms are ignored.
In \Eq.{DisInt} $({\rm Cont.})$ denoes the continuum contribution coming from the last term of \Eq.{ImPart},
which can be taken into account by multiplying the term $p^{2(n-1)}\ln(-p^2)$ in the OPE by
\begin{eqnarray*}
C_n(\omega)&=&1-\left[\sum_{k=1}^n{1\over(k-1)!}\left({\omega^2\over\MBS}\right)^{n-1}\right]
\exp\left(-{\omega^2\over\MBS}\right),
\end{eqnarray*}
after Borel transformation.

Substituting \Eq.{OPE} and  \Eq.{DisInt} into the left-hand side and the right-hand side of \Eq.{BSR}, respectively, we obtain sum rules for the coupling constant,
\begin{eqnarray}\label{SReven}
&&-{2\MN\over\MBS}\left(g-{g'\MN+g\over2\MN^2}\MBS\right)
{\sqrt 2\lambda^2\over \MBS}\exp\left(-{\MN^2\over \MBS}\right)
\cr&=&{\sqrt{2}\over4\pi^2}\Bigg\{
-\MB^2C_2(\omega_\pi)\Bigg[{\<>{\qq}\over f_\pi}+(E_q+\MN-\omega_k)2f_\pi\Bigg]
-C_1(\omega_\pi)[(E_q+\MN-\omega_k)2f_\pi\delta^2]
\cr& &
-{1\over \MB^2}\Bigg[-{\pi^2\over6}{\<>{\qq}\<>{\GG}\over f_\pi}
+(E_q+\MN-\omega_k){\pi^2\over3}\left({8\over3}{\<>{\qq}^2\over f_\pi}
+f_\pi\<>{\GG}\right)
\cr& &
-\omega_k{\pi^2\over3}\Bigg(8{\<>{\qq}^2\over f_\pi}+f_\pi\<>{\GG}
+(E_q+\MN)16f_\pi\<>{\qq}\Bigg)\Bigg]
\cr& &+{1\over\MB^4}\Bigg[
-(E_q+\MN-\omega_k){\pi^2\over9}{m_0^2\<>{\bar qq}^2\over f_\pi} 
-\omega_k{\pi^2\over3}{m_0^2\<>{\bar qq}^2\over f_\pi} 
\cr& &
-\omega_k(E_q+\MN){\pi^2\over3}\Bigg(4f_\pi m_0^2\<>{\bar qq}
-{448+32\epsilon\over21}f_\pi\delta^2\<>{\qq}
-{1\over3}{\<>{\qq}\over f_\pi}\<>{\GG} \Bigg)\Bigg]
 \Bigg\},
\cr&\equiv&L_B[\Pi_{+\;\even}^{\rm OPE-Cont.}]
\end{eqnarray}
and
\begin{eqnarray}\label{SRodd}
&&-{2\MN\over\MBS}\left(g-{g'\over 2\MN}\MBS\right)
{\sqrt 2\lambda^2\over \MBS}\exp\left(-{\MN^2\over \MBS}\right)
\cr&=&{\sqrt{2}\over4\pi^2}\Bigg\{
-C_1(\omega_\pi)\Bigg[-(E_q+\MN){\<>{\qq}\over3f_\pi}
-\omega_k\left({\<>{\qq}\over f_\pi}+(E_q+\MN)2f_\pi\right) \Bigg]
\cr& &
-{1\over \MB^2}\Bigg[-(E_q+\MN)\left({16\pi^2\over3}f_\pi\<>{\qq}
+\omega_k{16\over9}f_\pi\delta^2\right) \Bigg]
\cr& &+{1\over \MB^4}\Bigg[
-(E_q+\MN){\pi^2\over3}\left(2f_\pi m_0^2\<>{\bar qq}
-{1\over18}{\<>{\qq}\<>{\GG}\over f_\pi}
-{104\over9}f_\pi\delta^2\<>{\qq}\right)
\cr& &
+\omega_k{\pi^2\over3}\left(-{1\over2}{\<>{\qq}\<>{\GG}\over f_\pi}
+(E_q+\MN)f_\pi\<>{\GG}\right)\Bigg]\Bigg\},
\cr&\equiv&L_B[\Pi_{+\;\odd}^{\rm OPE-Cont.}],
\end{eqnarray}
where we have used the parametrization, $g_s\<>{\mix}=m_0^2\<>{\bar qq}$.

From the sum rule for the vacuum-to-vacuum matrix element of the same operator, the strength $\lambda^2$ is given up to the dimension seven by
\begin{eqnarray}\label{SRNeven}
\MN{\lambda^2\over \MBS}\exp\left(-{\MN^2\over\MBS}\right)
&=&{1\over4\pi^2}\left[\MB^2C_2(\omega_0)(-\<>{\qq})
+{\pi^2\<>{\qq}\<>{\GG}\over6\MB^2}\right]\cr
&\equiv&L_B[\Pi_{0\;\even}^{\rm OPE-Cont.}],
\end{eqnarray}
and
\begin{eqnarray}\label{SRNodd}
{\lambda^2\over \MBS}\exp\left(-{\MN^2\over\MBS}\right)
&=&{1\over4\pi^4}\left[{\MBQ\over8}C_3(\omega_0)
+{C_1(\omega_0)\over8}\pi^2\langle{\alpha_s\over\pi}G^2\rangle
+{8\pi^4\langle\bar qq\rangle^2\over3\MBS}\right]\cr
&\equiv&L_B[\Pi_{0\;\odd}^{\rm OPE-Cont.}],
\end{eqnarray}
where $\omega_0$ is the effective continuum threshold.

Now, we eliminate $\lambda^2$ by dividing the vacuum-to-pion sum rule by the vacuum-to-vacuum sum rule.
Since we have even and odd equations for the vacuum-to-pion and the vacuum-to-vacuum correlation functions, respectively, we obtain the following four sum rules
\begin{eqnarray}\label{SR}\nonumber
&&g-{g+g'M\over2\MN^2}\MBS
=-{\sqrt 2\MBS\over4\MN}{L_B[\Pi_{+\;\even}^{\rm OPE-Cont.}]\over
L_B[\Pi_{0\;\even}^{\rm OPE-Cont.}]},
\\\nonumber
&&g-{g'\over2\MN}\MBS
=-{\sqrt 2\MBS\over4\MN}{L_B[\Pi_{+\;\odd}^{\rm OPE-Cont.}]\over
L_B[\Pi_{0\;\even}^{\rm OPE-Cont.}]},
\\\nonumber
&&g-{g+g'M\over2\MN^2}\MBS
=-{\sqrt 2\MBS\over4\MN^2}{L_B[\Pi_{+\;\even}^{\rm OPE-Cont.}]\over
L_B[\Pi_{0\;\odd}^{\rm OPE-Cont.}]},
\\
&&g-{g'\over2\MN}\MBS
=-{\sqrt 2\MBS\over4\MN}{L_B[\Pi_{+\;\odd}^{\rm OPE-Cont.}]\over
L_B[\Pi_{0\;\odd}^{\rm OPE-Cont.}]}.
\end{eqnarray}

The procedure of extracting the coupling constant goes as follows.
We regard the right-hand side of the each of the above equations as a function of the Borel mass squared, $\MBS$.
\Eq.{SR} means that if the sum rule works there is a region in $\MBS$ where the right-hand side is well approximated by a linear function in $\MBS$.
In other words we rewrite the above sum rules in the form,
\begin{eqnarray}\label{l_eq}
g=f(\MBS)-\MBS{d\over d \MBS}f(\MBS),
\end{eqnarray}
where $f(\MBS)$ denotes the right-hand side of the sum rules, \Eq.{SR}, and look for plateau in $\MBS$.
If we find a plateau, we conclude the value of the plateau to be the sum rule prediction.
In this procedure the effective continuum thresholds are determined so as to provide the most stable plateau.

We are now ready to discuss the numerical results of the sum rules.
In the sum rules, we use the following parameters which determine vacuum-to-vacuum and vacuum-to-pion matrix elements of the quark-gluon composite operators,
\begin{eqnarray}\label{Parameter}
&& \<>\qq=-(225{\rm MeV})^3, \qquad
\<>{\GG}=(330{\rm MeV})^4, \cr
&& m_0^2=1{\rm GeV}^2, \qquad
\delta^2=0.2{\rm GeV}^2, \qquad
\epsilon=0.5.
\end{eqnarray}

We first discuss the results in the chiral limit.
We searched for the effective continuum thresholds, $\omega_0$ and $\omega_\pi$, between $1.44$~GeV and $\infty$ for which the calculated coupling constant, $g$, has the most stable plateau as a fanction of the Borel mass squared, $\MB^2$, over the range 1~GeV$^2\le \MB^2\le$ 2~GeV$^2$. 
The optimum choices are ($\omega_0$, $\omega_\pi$)= (2.4~GeV, 2.0~GeV), (1.9~GeV, 1.6~GeV), (2.0~GeV, 1.44~GeV) and (1.9~GeV, 1.6~GeV) for the first, second, third and fourth sum rules, respectively. 
The calculated coupling constant is plotted as a function of the Borel mass squared in Fig.~1.
In all four sum rules stable plateaus are found with the value between 7 and 9.5. 
The OPE of the second sum rule is the same as that for the Dirac structure $\gamma_5\sigma_{\mu\nu}p^\mu k^\nu$ in Ref[\ref{KLO}],
while the physical parametrization of the former is different from that of the latter since the single pole term vanishes in the former but exists in the latter.
However, the prediction of the former, $g=9$, is numerically close to the latter since the coefficient of the single pole term is relatively small as is pointed out in Ref~[\ref{KLO}].
The OPE of the first sum rule is a combination of the $i\gamma_5$ and $i\gamma_5\gsl k$ structures and its prediction, $g=9.5$, is closer to the empirical value, $g_{\rm emp.}=13.4$, than other sum rules.
The third and the fourth sum rules do not correspond to sum rules in previous works, since the former uses the odd part of the vacuum-to-vacuum sum rule while the latter uses the even one.
The calculated coupling constants of the third and the fourth sum rules are smaller than those of the first and the second sum rules.


We turn to the results including $O(\mpi)$ corrections.
We first examine if the effect of the contninuum is regarded as correction or qualitative change in four sum rules.
Figs.~2 and 3 show $g$ vs. Borel mass squared for $\omega_0=\omega_\pi=$ 1.44~GeV and $\omega_0=\omega_\pi=\infty$, respectively.
From Figs.~2 and 3 one sees that the third sum rule is very sensitive to the effective threshold while other three depend rather moderately on the effective threshold.
Thus, we exclude the third sum rule from the following analysis.
Then, as in the chiral limit we searched for the effective continuum thresholds, $\omega_0$ and $\omega_\pi$, between $1.44$~GeV and $\infty$ for which the calculated coupling constant, $g$, has the most stable plateau as a fanction of the Borel mass squared, $\MB^2$, over the range 1~GeV$^2\le \MB^2\le$ 2~GeV$^2$. The calculated coupling constants in the three sum rules are plotted in Fig.~4.
From Figs.~1 and 4 one sees that $O(\mpi)$ correction is large for the fourth sum rule which seems to indicate that the sum rule is unreliable.
In the second sum rule the coupling constant changes almost 40\% from $\MBS$=1~GeV$^2$ to 2~GeV$^2$.
Thus, the second sum rule does not satisfy the Borel stability.
In contrast the first sum rule has a very stable plateau.
Therefore, we conclude that the first sum rule is the best for extracting the $\pi NN$ coupling constant.
Comparing the solid line in Fig.~4 and in Fig.~1 we find the $O(\mpi)$ correction for the coupling constant in the first sum rule is about 5\%.

Finally, we disscuss the errors due to uncertainties of the parameters.
We found that the coupling constant is most sensitive to the quark condensate, typically quoted error, $\langle\bar qq\rangle=-(225\pm25{\rm MeV})^3$, changes
$g$ by $\pm1.7$.
The errors of $\<>{\GG}$ are expected to be irrelevant since the contribution from the gluon condensate is well known to be small in the nucleon sum rule.
We next study the errors due to $m_0^2$ and $\delta^2$.
The parameter, $m_0^2$, lies in the range between $0.6{\rm GeV}^2$ and $1.4{\rm GeV}^2$ according to Ref.~[\ref{m02}].
The numerical estimation, $\delta^2=0.2\pm0.02{\rm GeV}^2$, was performed by Novikov et al. through the QCD sum rules in Ref.~[\ref{Novikov}].
The errors due to uncertainty of $m_0^2$ and $\delta^2$ turn out to be less than 1\% and 5\%, respectively.
We expect that the uncertainty of $\epsilon$ is also unimportant, since from \Eq.{SReven} its contribution is too small.
We finally investigate the errors due to the uncertainties of the effective continuum thresholds.
From the Borel stability analysis we determined the errors of the effective continuum thresholds as $\omega_0=2.5\pm0.14$~GeV and $\omega_\pi=2.0\pm0.14$~GeV.These errors change $g$ by $\pm1.3$ and $\pm1.4$, respectively.
Taking account of all these uncertainties, we conclude
\begin{eqnarray}
g=10\pm3,
\end{eqnarray}
which is in reasonable agreement with the emprical value.

A comment for the choice of $\delta^2$ is in order here.
Birse and Krippa estimated the uncertainty of $\delta^2$ as $0.2{\rm GeV}^2\le\delta^2\le0.45{\rm GeV}^2$ and took $\delta^2=0.35{\rm GeV}^2$ in the calculation of $g$~[\ref{B&K}]. 
If we take $\delta^2=0.35{\rm GeV}^2$, from \Eq.{l_eq} we obtain
\begin{eqnarray}
g\approx14.
\end{eqnarray}
Thus the prediction of our sum rule does not become bad even if we take $\delta^2$ much larger than the typical value $0.2{\rm GeV}^2$.


\section{SUMMARY}
We have proposed a new approach in order to construct QCD sum rules for the $\pi NN$ coupling constant, $g$, starting from the vacuum-to-pion correlation function of the interpolating fields of two nucleons and taking its matrix element with respect to nucleon spinors.
This is based on the detailed study of the physical content of the correlation function without referring to the effective lagrangian.
In the chiral limit either the double-pole term vanishes or the single-pole term as well as the continuum term is not separately well defined under the dispersion integral in previous approaches using invariant correlation functions, while the double-pole term survives but the single-pole term and the continuum term give well-defined contributions in our new approach with projected correlation functions.
In this respect the use of the projected correlation function seems advantageous.
Another advantage of our approach is that unwanted pole contribution is projected out.

Calculating the Wilson coefficients of the OPE of the correlation function up to $O\left(\MB^{-4}\right)$ and $O\left(\mpi\right)$ where $\MB$ and $\mpi$ are the Borel mass and the pion mass, respectively,
we have constructed new QCD sum rules for the $\pi NN$ coupling constant from the projected correlation function with consistently including $O(\mpi)$ corrections.
Since there are even and odd terms both for the vacuum-to-pion and the vacuum-to-vacuum sum rules, by dividing former by the latter in order to eliminate the coupling strength of the interpolating field to the nucleon we have derived four sum rules.

We examined if the pole contribution dominates the continuum contribution, if $O(\mpi)$ correction is reasonably small and if the result is stable in the sense of Borel stability.
We concluded that among four sum rules the one obtained by dividing the even vacuum-to-pion sum rule by the even vacuum-to-vacuum sum rule is most pertinent.
After roughly estimating errors we obtaind, $g=10\pm3$, as a result of this sum rule, which is in reasonable agreement with the empirical value.
.
We also found that  the $O(\mpi)$ correction is about 5\%.

The present approach can be used for other interesting applications.
First of all, if we keep the pion momentum finite in the present formalism, we will obtain the information on the $\pi N N$ form factors.
It is also possible to calculate not only the coupling constants of meson and two same baryons like $\eta NN$ and $\pi\Sigma\Sigma$ but also the coupling constants of meson and two different baryons like $K N\Sigma$ and $\pi\Lambda\Sigma$ without taking the flavor SU(3) limit.
\\

\acknowledgments

We would like to thank Makoto Oka and Takumi Doi for helpful discussions and valuable comments.

\newpage

\baselineskip 24pt


\def\labelenumi{[\theenumi]}
\def\Ref#1{[{#1}]}
\def\npb#1#2#3{{Nucl. Phys.\,} {\bf B{#1}}\,(#3)\,#2}
\def\npa#1#2#3{{Nucl. Phys.\,} {\bf A{#1}}\,(#3)\,#2}
\def\np#1#2#3{{Nucl. Phys.\,} {\bf{#1}}\,(#3)\,#2}
\def\plb#1#2#3{{Phys. Lett.\,} {\bf B{#1}}\,(#3)\,#2}
\def\prl#1#2#3{{Phys. Rev. Lett.\,} {\bf{#1}}\,(#3)\,#2}
\def\prd#1#2#3{{Phys. Rev.\,} {\bf D{#1}}\,(#3)\,#2}
\def\prc#1#2#3{{Phys. Rev.\,} {\bf C{#1}}\,(#3)\,#2}
\def\pr#1#2#3{{Phys. Rev.\,} {\bf{#1}}\,(#3)\,#2}
\def\ap#1#2#3{{Ann. Phys.\,} {\bf{#1}}\,(#3)\,#2}
\def\prep#1#2#3{{Phys. Rep.\,} {\bf{#1}}\,(#3)\,#2}
\def\rmp#1#2#3{{Rev. Mod. Phys.\,} {\bf{#1}}\,(#3)\,#2}
\def\cmp#1#2#3{{Comm. Math. Phys.\,} {\bf{#1}}\,(#3)\,#2}
\def\ptp#1#2#3{{Prog. Theor. Phys.\,} {\bf{#1}}\,(#3)\,#2}
\def\ib#1#2#3{{\it ibid.\,} {\bf{#1}}\,(#3)\,#2}
\def\zsc#1#2#3{{Z. Phys. \,} {\bf C{#1}}\,(#3)\,#2}
\def\zsa#1#2#3{{Z. Phys. \,} {\bf A{#1}}\,(#3)\,#2}
\def\intj#1#2#3{{Int. J. Mod. Phys.\,} {\bf A{#1}}\,(#3)\,#2}
\def\sjnp#1#2#3{{Sov. J. Nucl. Phys.\,} {\bf #1}\,(#3)\,#2}
\def\pan#1#2#3{{Phys. Atom. Nucl.\,} {\bf #1}\,(#3)\,#2}

\def\etal{{\it et al.}}
\begin{enumerate}

\divide\baselineskip by 4
\multiply\baselineskip by 3

\item \label{SVZ} M. A. Shifman, A. I. Vainshtein and V. I. Zakharov, 
\npb{147}{385}{1979}; \npb{147}{448}{1979}.
\item \label{RRY} L. J. Reinders, H. R. Rubinstein and S. Yazaki,
\prep{127}{1}{1985}.
\item \label{RRY2} L. J. Reinders, H. R. Rubinstein and S. Yazaki, 
\npb{213}{109}{1983}.
\item \label{S&H} H.~Shiomi and T.~Hatsuda, 
\npa{594}{294}{1995}.
\item \label{B&K} M.~C.~Birse and B.~Krippa, 
\plb{373}{9}{1996}; \prc{54}{3240}{1996}.
\item \label{KLO} H. Kim, S.H. Lee and M. Oka,
\plb{453}{199}{1999};\\ \prd{60}{034007}{1999}.
\item \label{DKO} T. Doi, H. Kim and M. Oka,
\prc{62}{055202}{2000}.
\item \label{Maltman} K~Maltman, 
\prc{57}{69}{1998}.
\item \label{recentKM} Y. Kondo and O. Morimatsu,
\prc{66}{028201}{2002}.
\item \label{K&M} Y. Kondo and O. Morimatsu,
\ptp{100}{1}{1998}.
\item \label{Ioffe} B. L. Ioffe,
\npb{188}{317}{1981}; \npb{191}{591(E)}{1981}.
\item \label{Belyaev} 
V. M. Belyaev, V. M. Braun, A. Khodjamirian and R. R\"uckel,\\
\prd{51}{6177}{1995}.
\item \label{Braun}
V. M. Braun, I. B. Filyanov, \zsc{48}{239}{1990}.
\item \label{m02} A.~A.~Ovchinnikov and A.~A.~Pivovarov,\sjnp{48}{721}{1988}; 
M.~Kremer and G.~Schierholz, \plb{194}{283}{1987};
M.~V.~Polyyakov and C.~Weiss, \plb{387}{841}{1996}.
\item \label{Novikov} V.~A.~Novikov, M. A. Shifman, A. I. Vainshtein, M.~B.~Voloshin and V.~I.~Zakharov, \npb{237}{525}{1984}.

\end{enumerate}


\newpage

\begin{figure}
  \begin{center}
    \leavevmode
    \psfig{figure=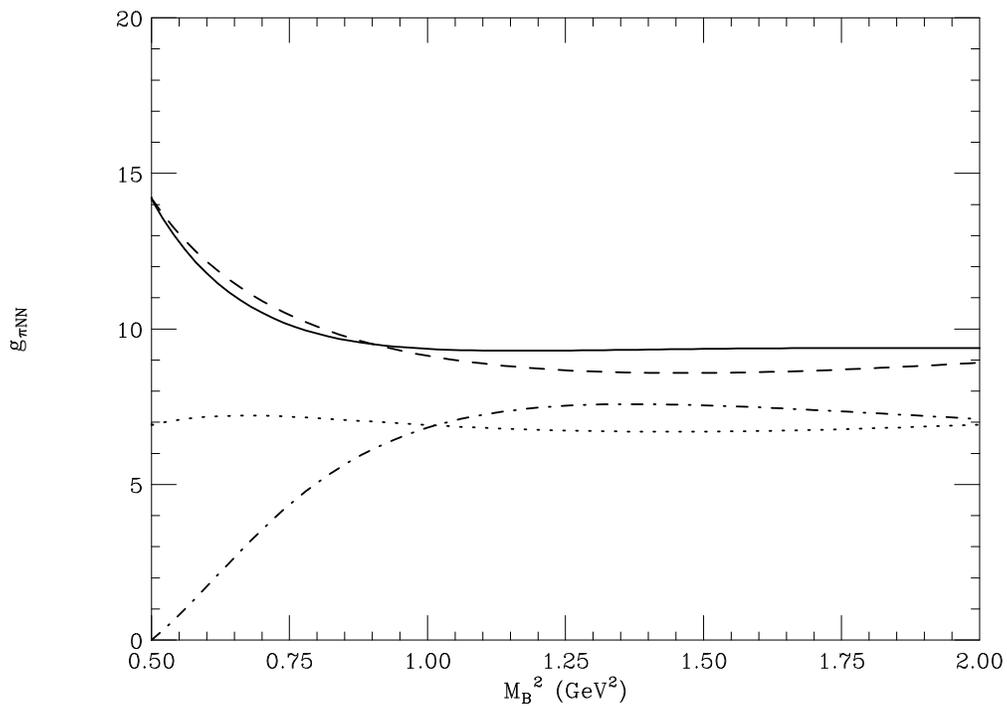,angle=90,width=5.2in}
  \end{center}
  \caption{The calculated $\pi NN$ coupling constant in the chiral limit with $k=0$ vs. the Borel mass squared, $\MBS$.
The first, second, third and fourth sum rules in Eq.~(29) are for solid, dashed, dot-dashed and dotted lines, respectively.
The effective continuum thresholds, $(\omega_0,\omega_\pi)$, determined by the Borel stability analysis are $(2.4{\rm GeV},2.0{\rm GeV})$, 
$(1.9{\rm GeV},1.6{\rm GeV})$, 
$(2.0{\rm GeV},1.44{\rm GeV})$ 
and $(1.9{\rm GeV},1.6{\rm GeV})$
for solid, dashed, dot-dashed and dotted lines, respectively.}
  \label{fig:1}
\end{figure}

\newpage

\begin{figure}
  \begin{center}
    \leavevmode
    \psfig{figure=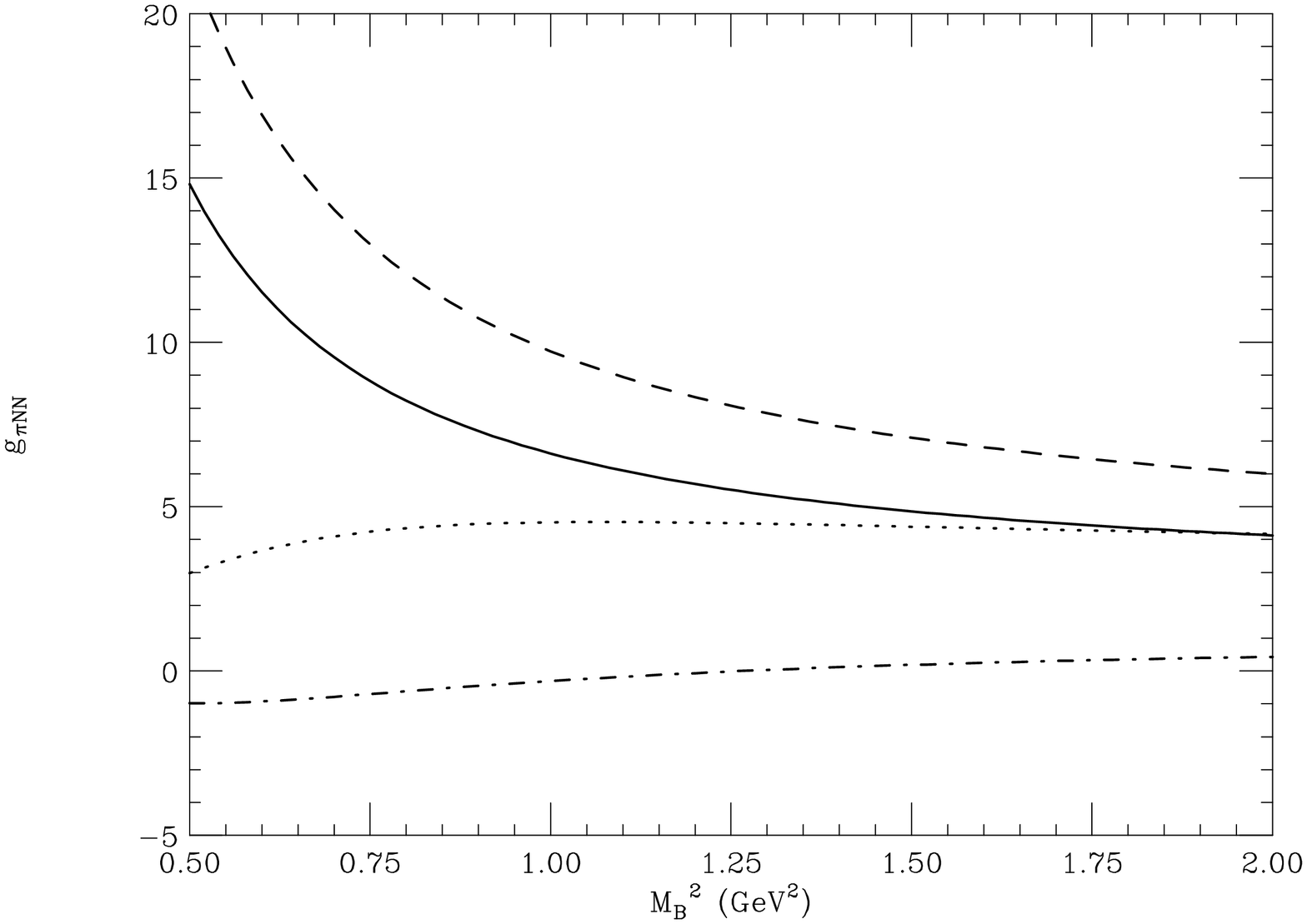,angle=90,width=5.2in}
  \end{center}
  \caption{The calculated $\pi NN$ coupling constant vs. the Borel mass squared, $\MBS$. The effective continuum thresholds are $\omega_0=\omega_\pi=1.44{\rm GeV}$ for all lines. 
The first, second, third and fourth sum rules in Eq.~(29) are for solid, dashed, dot-dashed and dotted lines, respectively.}
  \label{fig:2}
\end{figure}

\newpage

\begin{figure}
  \begin{center}
    \leavevmode
    \psfig{figure=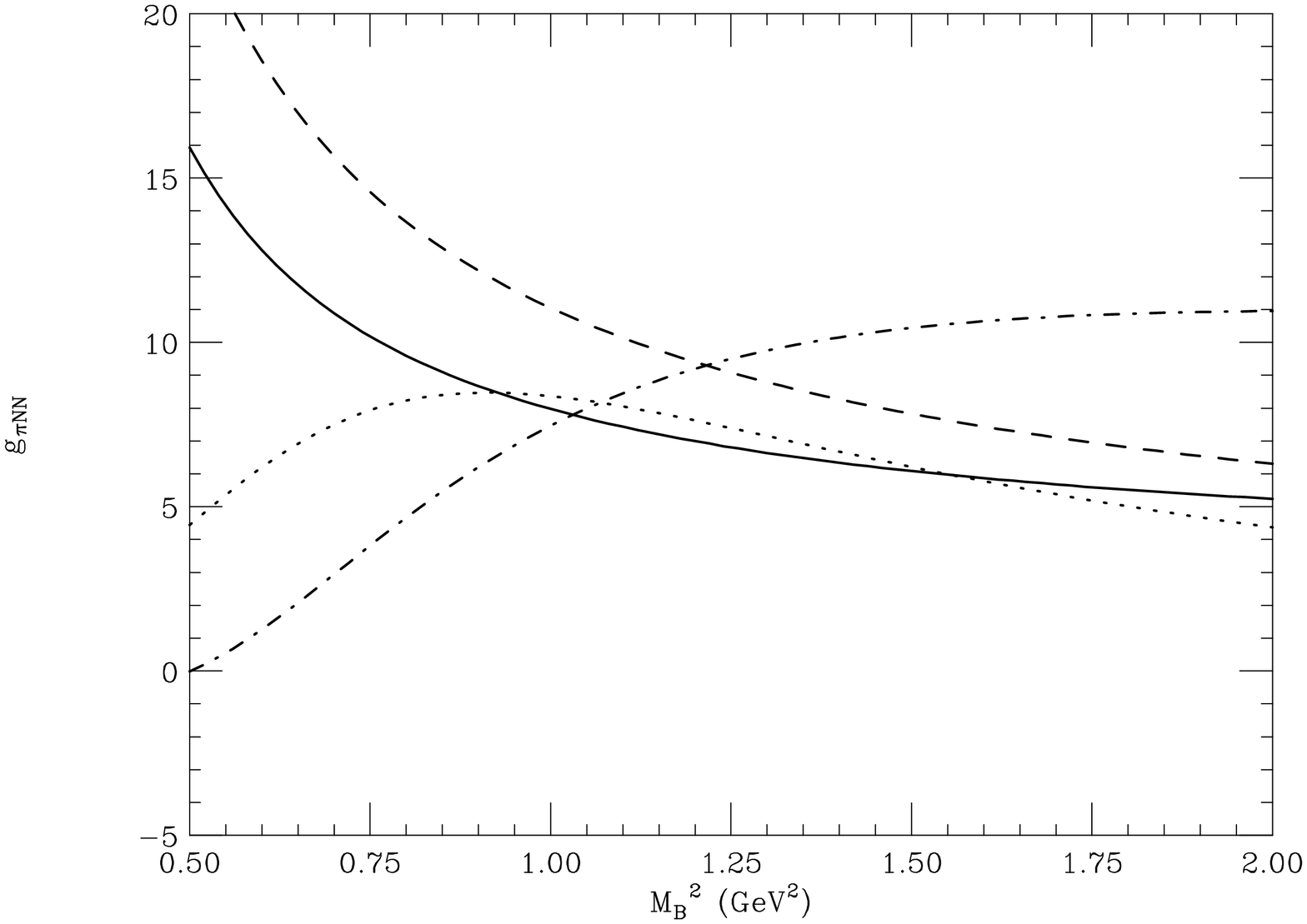,angle=90,width=5.2in}
  \end{center}
  \caption{The calculated $\pi NN$ coupling constant vs. the Borel mass squared, $\MBS$. The effective continuum thresholds are $\omega_0=\omega_\pi=\infty$ for all lines. 
The first, second, third and fourth sum rules in Eq.~(29) are for solid, dashed, dot-dashed and dotted lines, respectively.}
  \label{fig:3}
\end{figure}

\newpage

\begin{figure}
  \begin{center}
    \leavevmode
    \psfig{figure=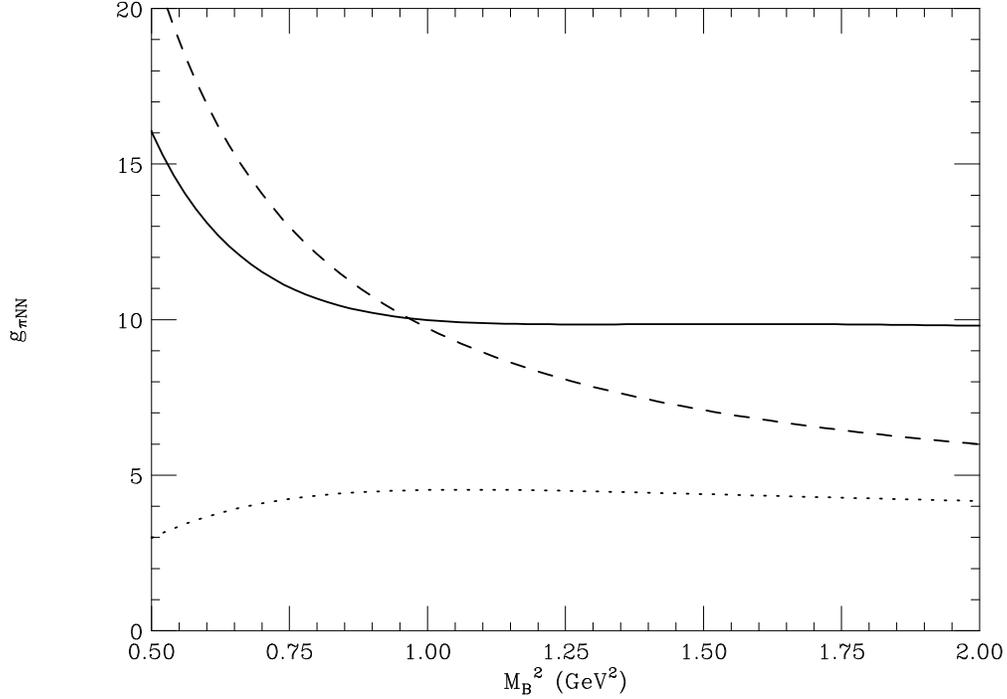,angle=90,width=5.2in}
  \end{center}
  \caption{The calculated $\pi NN$ coupling constant vs. the Borel mass squared, $\MBS$. 
The first, second and fourth sum rules in Eq.~(29) are for solid, dashed and dotted lines, respectively.
The effective continuum thresholds, $(\omega_0,\omega_\pi)$, determined by the Borel stability analysis are $(2.5{\rm GeV},2.0{\rm GeV})$, 
$(1.44{\rm GeV},1.44{\rm GeV})$ 
and $(1.44{\rm GeV},1.44{\rm GeV})$
for solid, dashed and dotted lines, respectively.
}
  \label{fig:4}
\end{figure}

\end{document}